\documentstyle[sprocl]{article}

\def\be{\begin{equation}}
\def\ee{\end{equation}}
\def\bea{\begin{eqnarray}}
\def\eea{\end{eqnarray}}
\def\sl#1{\rlap/#1}

\begin{document}
\renewcommand{\thefootnote}{\fnsymbol{footnote}}
\title{Induced Lorentz-Violating Chern-Simons Term in QED:
Uncovering Short Distance Interaction Terms in the Effective Lagrangian
without the Shadow of Regularization\footnote{Talk presented at the
Second Meeting on CPT and Lorentz Symmetry CPT'01, August 15-18,
2001, Bloomington, Indiana.}}

\author{Lai-Him Chan}
 
\address{Department of Physics and Astronomy,
       	Louisiana State University,
	       Baton Rouge, Louisiana 70803-4001, USA\\E-mail:chan@phys.lsu.edu}


\maketitle\abstracts{ We show that the correctly evaluated effective
Lagrangian should include short-distance interaction terms which have
been avoided under the protection of usual regularization and must be
properly identified and reinstated if regularization is to be
removed. They have special physical and mathematical significance as
well as restoring gauge invariance and suppressing divergence in the
effective Lagrangian.  The rich structure of the short-distance
interaction terms can open up challenging opportunities where the
conventional regularization with rigid structure is unavailable and
inappropriate. It becomes clear that gauge invariance is preserved
with or without regularization and therefore there is no
Lorentz-Violating Chern-Simons term in QED. }

There has been a serious controversy generated by a series of recent
papers whether a Lorentz-violating CPT-odd term in a charged fermion
Lagrangian,
\begin{eqnarray}\label{Lb}
{\cal L}=\bar{\psi}\Big[i\sl{\partial}-m-
Q\,\sl{\!A(x)}-\gamma_5 \sl{b}\Big]\psi\ ,
\end{eqnarray}
can induce a finite Lorentz-violating term 
${1\over2}\epsilon_{\mu\alpha\beta\nu}k^\mu 
F^{\alpha\beta}A^\nu$ to the
Lagrangian in the electromagnetic sector. The values of $k_\mu$ from
theoretical calculations varies from $0$, ${3\over8\pi^2}Q^2 b_\mu$,
${1\over8\pi^2}Q^2 b_\mu$ to completely
undetermined  \cite{cg2,jk,PVChung,Jk,chep,chen}.

The effective action resulted from integrating out the fermion field
in Eq.(\ref{Lb}) is given by,
\begin{equation}\label{Actionb}
{\Gamma}_{eff}=-i\,Tr\,ln\Big[i\sl{\partial}-Q\,\sl{\!A(x)}-m 
-\gamma_5\sl{b}\Big]\ ,
\end{equation}
The $T\!r$ here denotes both the trace $tr_\gamma$ in the
Dirac spinor space and the trace $tr_x$ in the coordinate space.
Using the following method to evaluate $tr_x$ \cite{ccz},
\begin{eqnarray}\label{trx}
\!\!tr_{\!x}&&\!\!\!\!f(i\partial\!-\!QA(x))\!=\!\!\!\int
\!\!\!d^{4}\!x\!\!\lim_{x'\to x}\!\langle x'|f(i\partial\!-
\!QA(x))|x\rangle\!\!=\!\!\!\int\!\!\!d^{4}\!x\!\lim_{x'\to
x}\!\!f(i\partial\!-\!QA(x))\delta(x\!-\!x')\nonumber\\ 
=&&\!\!\int\!\!\!d^{4}x\!\!\int
\!\!\!{d^4 p\over(2\pi)^4}\!\!\lim_{x'\to
x}\!f(i\partial-QA(x))e^{-ip.(x-x')}\!\!
=\!\!\!\!\int \!\!d^{\!4}x\!\!\!\int
\!\!\!{d^4 p\over(2\pi)^4}f(i\partial\!+\!p\!-\!QA(x)),
\end{eqnarray}
we obtain the effective Lagrangian in a compact form from 
Eq.(\ref{Actionb})
\begin{equation}\label{Leffb}
{\cal L}_{eff}=-i\int\!\!{d^4
p\over(2\pi)^4}tr_\gamma\,ln\,
G^{-1}(p+i\partial,x),
\end{equation}
where $G(p+i\partial,x)=\left[\sl{p}-m+i\sl{\partial}-Q\,\sl{\!A(x)}
-\gamma_5\sl{b}\right]^{-1}$ and $p$ is the fermion loop momentum.

When  gauge invariant regularization is used, the effective Lagrangian
is gauge invariant and  naturally $k_\mu=0$ \cite{cg2,Bo}. It has been
argued that if the the gauge noninvariant anomalous contributions is
finite  without regularization, then the role of a gauge-invariant
regularization is no longer regulating the divergent quantities  but
rather imposing gauge invariant \cite{jk,Jk,chep}.  The option of
removing  gauge-invariant regularization should be kept open, in this
case as in the chiral anomaly calculation. However, without the
protection of regularization, it is not clear to what degree
short-distance dynamics has been or can ever be properly defined.
Therefore $k_\mu$ may not be determined \cite{chen}.

To uncover possible short-distance terms, we have to make precise
calculation of the effective Lagrangian Eq.(\ref{Leffb}) \cite{chep}.
Without the protection of regularization, the intrinsic quantum
incompatibility, the noncommutativity between the photon fields  and
the space-time derivative, becomes relevant \cite{chep}. The power
expansion of the photon fields and the power expansion of the
derivative (small momentum) expansion become incompatible when the
coefficients of lower power of $A$ become singular.  This difficulty
can be only circumvented by expanding the effective Lagrangian in
powers of
$\sl{\Pi}=i\sl{\partial}-Q\,\sl{\!A(x)}$ and $\gamma_5\sl{b}$. We obtain 
the Chern-Simons Lagrangian \cite{chep},
\begin{eqnarray}
&{\cal L}_{eff}^{CS}&=4i\,\epsilon_{\mu\alpha\beta\nu}b^\mu
\Pi^{\alpha}\Pi^{\beta}\Pi^\nu\!
\int\!\! \!{d^4
p\over(2\pi)^4}{m^2\over (p^2-m^2)^3}={i\over16\pi^2}Q^2
F^{\alpha\beta} A^\kappa
\epsilon_{\lambda\alpha\beta\nu}\nonumber.
\end{eqnarray}
In order to compare our result with the calculation basing on the
Feynman diagram approach, we compute the vacuum polarization tensor,
\begin{eqnarray}
\Pi^{\mu\nu}(q)&=&\int\!{d^4 p\over(2\pi)^4}
tr_\gamma\,\Bigg\{\gamma^\mu{i\over
\sl{p}-m-\gamma_5\sl{b}}\gamma^\nu{i\over
\sl{p}+\sl{q}-m-\gamma_5\sl{b}}\label{vacp}\\
&+& i \,q_\alpha{\partial\over\partial p^\alpha}\int^\infty_m\!\!\!\!
dM\Big[\gamma^\mu\Big({i\over\sl{p}\!-\!M\!-\!\gamma_5\sl{b}}
\Big)^{\!\!2}
\gamma^\nu {i\over\sl{p}\!+\!\sl{q}\!-\!M\!-\!\gamma_5\sl{b}}
\Big]\!\Bigg\}\nonumber
\end{eqnarray}
which is written in terms of the standard one-loop vacuum polarization
Feynman amplitude with
$b_\mu$-exact fermion propagators
\cite{jk} plus a $b_\mu$-exact correction from the
noncommutativity of the operators $A$ and $\partial$ \cite{chep}. This
correction gives precisely the additional contribution 
$-{1\over32\pi^2}Q^2b_\mu $ to $k_\mu$ to be combined with the
result from previous calculation \cite{jk} of the first term
to yield
$k_\mu={3\over32\pi^2}Q^2b_\mu-{1\over32\pi^2}Q^2b_\mu=
{1\over8\pi^2}Q^2b_\mu$.
\goodbreak

By removing regularization, gauge dependent short-distance interaction
terms, in the form of momentum integral of total momentum derivative 
not expressible in term of conventional Feynman diagrams, are exposed
and  can make finite contributions to the effective Lagrangian. The
finite nongauge-invariant anomalous contributions are due to the
short-distance singularity and sensitively depend on the path how $x'$
approaches to $x$. In Eq.(\ref{trx}),  $tr_x$ should be
performed symmetrically with respect to $x$ and $x'$, However, the
reduction from Eq.(\ref{Actionb}) to Eq.(\ref{Leffb}) by
Eq.(\ref{trx}) is explicitly asymmetric in $x$ and
$x'$ because the operator $i\partial-QA(x)$ is not diagonal in
$x$ and cannot be diagonalized. It appears that there is no simple way
to escape from regularization. However,  there is a very good reason
and a solution for this problem. 

In the semiclassical approach, the particle field is divided into a
classical part which satisfies the classical field equation and the
quantum part which obeys the quantum canonical commutation or anticommutation
relations.  In the path integral approach, the effective action is
obtained by functionally integrating out the quantum fluctuation due
to the quantum part. Likewise, in the evaluation of $tr_x$,  the pair
of space-time canonical operators in Eq.(\ref{Leffb}),
$\{p+i\partial,x\}$,  is the sum of the quantum part  and the classical
part.   The pair $\{i\partial,x\}$
obeying the same canonical commutation relation is obviously the
quantum part  and $\{p,0\}$  diagonal in $p$ must be the classical part.
Such decomposition is completely opposite to the original semiclassical
approach that one would expect. Namely,
$x$, being the space-time coordinates of the background field, should
be classic and the loop-momentum $p$ 
representing the integration of quantum fluctuation should be the
quantum part. Therefore the decomposition of the canonical pair in
Eq.(\ref{Leffb}) should have been
$\{p,x-i{\partial\over\partial p}\}$. 
The two decompositions are
related by a gauge covariant unitary transformation
$e^{\Pi\cdot{\partial\over\partial p}}$ in the direct product space of
$x$ and $p$ \cite{MaryK}. The expansion to this larger space allows the
diagonalization of $x$.

Therefore the correct effective Lagrangian  with interaction $\chi(x)$
in any dimension
$N$ should be,
\begin{eqnarray}
{\cal L}_{eff}\label{LGN}
&=&-i\,\int {d^Np\over(2\pi)^N}\,
tr_\gamma\,\left[e^{-\Pi\cdot{\partial\over\partial p}}\right] ln\,
(\sl{p}+\sl{\Pi}-m-\chi(x))
\left[e^{\Pi\cdot{\partial\over\partial p}}\right] ,
\end{eqnarray}
which is gauge invariant and diagonal in $x$ but
not in $p$. The corresponding propagator is,
\begin{eqnarray}\label{Greenfunction}
G(p,\Pi)=\left[e^{-\Pi\cdot{\partial\over\partial p}}\right]{i\over
\sl{p}+\sl{\Pi}-m-\chi(x)}\left[e^{\Pi\cdot{\partial\over\partial p}}
\right].
\end{eqnarray}
In Eq.(\ref{LGN}), the expansion of $e^{\Pi\cdot{\partial\over\partial
p}}$ gives no contribution. The expansion of
$e^{-\Pi\cdot{\partial\over\partial p}}$  yields  correction terms
proportional to the  total derivative of  $p$. They have been safely
discarded in conventional calculation with regularization. They must be
reinstated if regularization is not used.  Those correction terms can
be interpreted as short-distance gauge interaction and they are just as
well defined as conventional minimal gauge interaction in the
Lagrangian by the requirement for maintaining gauge invariance and
minimizing of the degree of divergence. Without regularization, it
takes the coordination of both the long distance interaction and short
distance interactions to fulfill that requirement.  There is also a
simple  physical interpretation. Photon emission is accompanied not
only with the change of momentum of charge classical momentum but also
with the change of loop momentum due the quantum fluctuation at short
distance.
The Green's function in Eq.(\ref{Greenfunction}) is similar in form to
the Schwinger's point splitting method of bilinear current. Here, the
short distance singularity due to two field operators at the same
space-time point is averaged out due to quantum fluctuation rather than
classical average.
$e^{(i\partial-eA)\cdot{\partial\over\partial p}}$ is a covariant 
displacement operator
to displace $x$ by $i {\partial\over\partial p}$.
\begin{eqnarray}
e^{(i\partial-eA)\cdot{\partial\over\partial
p}}\psi(x)=\psi(x+i{\partial\over\partial p})
\end{eqnarray}
Such operator can easily be constructed for any gauge symmetry and
any  particle fields.

We shall use the schwinger model, the $1\!+\!1$ dimensional massless
QED, to test the validity of the short-distance interaction terms in the
effective Lagrangian. We rewrite Eq.(\ref{LGN}) with $N\!=\!2$,
$m\!=\!0$, and
$\chi(x)\!=\!0$ in a form ready to expand in power series of
$A$.
\begin{eqnarray}\label{Lb2}
{\cal{L}}_{eff}\!=-\! \int\!\!{d^2p\over(2\pi)^{2}}
\,\left[e^{-(i\partial-eA)\cdot{\partial\over\partial p}}\right]
\!\int_0^\infty\!\!dm\, 
tr_\gamma{1\over\sl{p}+i\sl{\partial}-\!e\sl{A}\!-\!m}
.
\end{eqnarray} 
We have dropped the factor $e^{\Pi\cdot{\partial\over\partial
p}}$ because it has no contribution. The vacuum
polarization can be calculated directly without any regularization,
\begin{eqnarray}
\Pi^{\mu\nu}(p)\!&=&
{e^2\over\pi}\left({1\over2}g_{\mu\nu}-{p_\mu p_\nu\over p^2}\right)
-{e^2\over2\pi}g_{\mu\nu}+{e^2\over\pi}g_{\mu\nu}=
 {e^2\over\pi}\left(g_{\mu\nu}-{p_\mu p_\nu\over p^2}\right).
\end{eqnarray}
The first term corresponds to the contribution from conventional
calculation without the exponential correction factor \cite{Jk}. The
second term comes purely from the short distance correction factor. The
last term is from the mixed contribution. The net result recovers the
correct solution with photon mass $M^2={e^2\over\pi}$. Therefore the
short distant interaction terms from the gauge
covariant unitary transformation has indeed fulfilled its promised
role of restoring gauge invariance as well as contributing the
essential correction to the unregulated vacuum polarization.  

The CPT and Lorentz-violating effective Lagrangian corresponds to
the Eq.(\ref{LGN}) with $N=4$ and $\chi(x)=\gamma_5\sl{b}$. Since
the short-distance interaction terms  has restored gauge invariance, 
It is obvious that there is no induced Lorentz-Violating
Chern-Simons term.

We can now safely abolish the regularization barrier and 
reclaim the territory of short distance interaction in quantum
field theory. This new degree of freedom as viewed by individual
Feynman amplitudes appears to be chaotic. However in the
background field hybrid effective Lagrangian approach, we have the
benefit of viewing the generating functional of the complete set of
Feynman amplitudes and a new prospective begin to emerge. We have
demonstrated that not only short-distance interaction can be
derived to replace the effect of usual regularization but the
process can have real mathematical and physical significance. The
short-distance interaction can be considered as the quantum images of
the long-distance interaction in the classical Lagrangian. They must be
chosen in tandem to maintain gauge symmetries and to minimize the
degree of divergence in higher order perturbation calculations. It will
be a great challenge to apply this new tool to the cases where the
conventional regularization is unavailable or inappropriate.

\end{document}